\documentclass[12pt]{iopart}

\usepackage{epsf}

\newcommand{\Om}{$\Omega^-$}
\newcommand{\Mo}{$\overline{\Omega}^+$}
\newcommand{\X}{$\Xi^-$}
\newcommand{\Ix}{$\overline{\Xi}^+$}
\newcommand{\ptee}{$p_\mathrm{t}$}
\newcommand{\meanpt}{$\langle p_\mathrm{t}\rangle$}
\newcommand{\nineH}{$\sqrt{s}=900$ GeV}
\newcommand{\seven}{$\sqrt{s}=7$ TeV}
\newcommand{\twoH}{$\sqrt{s_{NN}}=200$ GeV}
\newcommand{\twosevensix}{$\sqrt{s_{NN}}=2.76$ TeV}
\newcommand{\pp}{pp}

\begin{document}

\title[]{Multi-strange particle measurements in 7 TeV proton-proton and 2.76 TeV PbPb collisions with the ALICE experiment at the LHC}

\author{D.D. Chinellato$^{1}$ for the ALICE Collaboration}
\address{$^{1}$ Universidade Estadual de Campinas (UNICAMP), Campinas, Brazil }
\ead{daviddc@ifi.unicamp.br}

\begin{abstract}
The production of charged multi-strange particles is studied with the ALICE experiment at the CERN LHC. Measurements of the central rapidity yields of \X\ and \Om\ baryons, as well as their antiparticles, are presented as a function of transverse momentum (\ptee) for inelastic \pp\ collisions at \seven\ and compared to existing measurements performed at the same and/or at lower energies. The results are also compared to predictions from two different tunes of the PYTHIA event generator. We find that data significantly exceed the production rates from those models. Finally, we present the status of the multi-strange particle production studies in Pb-Pb at \twosevensix\ performed as a function of collision centrality.\end{abstract}

\section{Introduction}

The study of charged multi-strange particles in high-energy proton-proton collisions serves not only as an important benchmark for heavy-ion collisions, 
but it can also provide us with information pertinent to the interplay between different particle production mechanisms. At low transverse momenta, soft
interactions are the dominant source of particle production, while at high \ptee\ perturbative quantum chromodynamics (pQCD) processes take over as the most significant mechanism. 
Information on these mechanisms can be used to constrain the phenomenology of QCD-inspired models and event generators such as PYTHIA \cite{pythia6.4}. The multi-strange
baryons are of particular interest as the colliding particles have no net strangeness content, so that any open strangeness must have been created in the collision.

In heavy-ion collisions, multi-strange baryon measurements can reveal information about the early partonic stages of the event, since these baryons have small
hadronic cross-sections. In Au-Au systems at \twoH\ at RHIC, radial flow was observed for the $\Omega$ that must have been accumulated 
prior to chemical freezeout \cite{exthechal}. 

\section{Data Analysis}

For this work, we studied multi-strange baryon production using a sample of $130\times10^{6}$ pp events measured by the ALICE experiment in 2010. 
The cascade decay candidates at mid-rapidity ($|y|\leq0.5$) are reconstructed out of combinations of high-quality tracks as measured by the ALICE time projection
 chamber (TPC) and the silicon Inner Tracking System (ITS) \cite{ALICE_general}. Further selections applied to these candidates include topological 
cuts and particle identification by energy loss while traversing the TPC gas. In addition, for the $\Omega$, we reject all candidates whose invariant mass under the $\Xi$ mass 
hypothesis is less than 8 $\rm MeV/c^{2}$ from the $\Xi$ nominal mass. 

The number of counts in each \ptee\ bin is taken within $4.5\sigma$ from
the invariant mass peak after background subtraction, where we determine the background in the regions from $4.5\sigma$ up to $9\sigma$ away from the peak. 
In order to compute detection efficiency, we used $\Xi^{\pm}$ or $\Omega^{\pm}$ enhanced Monte Carlo event generators such that every simulated event 
has at least one $\Xi$ or $\Omega$ baryon, and subsequent propagation of the particles is performed through the full ALICE geometry with GEANT3 \cite{Geant3Ref,Geant3Ref2}. 

Systematic errors are deduced point-by-point for the topological selections, signal extraction and track quality cuts; the \ptee-independent 
uncertainties contributing to the systematic error are normalization, material budget, GEANT3 (anti-)proton cross-section, TPC particle identification
 and mass under the $\Xi^{\pm}$ hypothesis rejection for the $\Omega^{\pm}$. 

\section{Results and Discussion}

\begin{figure}
\centerline{\epsfxsize 3.35in \epsffile{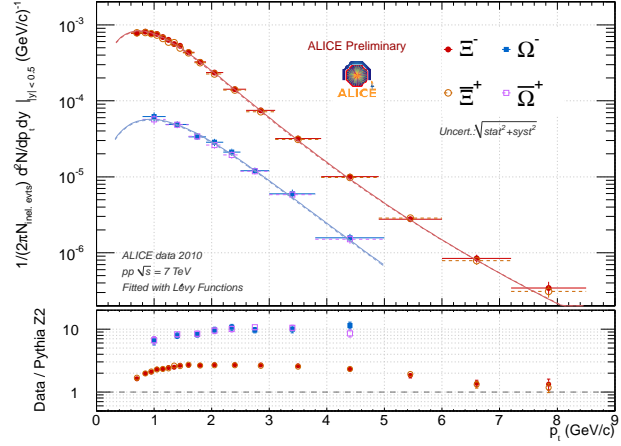}}
\caption{\label{ActualSpectra}  \X, \Ix, \Om\ and \Mo\ transverse momentum spectra. Also shown are ratios to Pythia Z2 predictions.}
\end{figure}

The $\Xi^{\pm}$ and $\Omega^{\pm}$ efficiency-corrected spectra are shown in Fig. \ref{ActualSpectra}. The anti-particle to particle ratios are consistent
with unity throughout the measured \ptee\ region for both multi-strange baryon species. We fit the spectra with L\'{e}vy functions as 
done previously in the ALICE \nineH\ \pp\ strangeness analyses \cite{ALICE_str_pp}. An additional 8$\%$ normalization uncertainty is not shown in Fig. \ref{ActualSpectra}.

The wide range in observed transverse momentum leads to an extrapolated area of only $22\%$ for the $\Xi^{\pm}$ and $26\%$ for the $\Omega^{\pm}$. The $\Xi^{\pm}$ integrated yield is approximately $60\%$ higher
than the one measured by ALICE at \nineH\ for inelastic events \cite{ALICE_str_pp}. The mean \ptee\ of $\Omega^{\pm}$ is about $20\%$ higher than for $\Xi^{\pm}$. 

When compared to Monte Carlo event generators, the total yields within the inelastic event category are $2.5\pm0.3$ times 
higher than the PYTHIA 6.4 Perugia-0 \cite{PerugiaTunesReference} predictions for $\Xi^{\pm}$ and $8.0\pm1.1$ times higher for $\Omega^{\pm}$. Newer tunes have been altered to match charged particle production measurements from \pp\ 
collisions at 7 TeV, such as the Z2 tune \cite{newtunes}; however, the Z2 tune still underpredicts total $\Xi^{\pm}$ production by a factor of $1.9\pm0.2$ and 
$\Omega^{\pm}$ production by a factor of $6.0\pm0.8$. In either tune used, the shape of the \ptee\ spectra is not well described, as shown in Fig. \ref{ActualSpectra} for the Z2 tune, 
and the \meanpt\ is larger in data by about $20\%$. The one notable exception is the $\Xi^{\pm}$ spectrum at high \ptee, where the predictions and the data seem to converge. 

When compared to other experiments, the measured yield for the $\Xi^{\pm}$ is smaller than the one measured by CMS \cite{CMS_str_pp}. This is due to
the difference in normalization, as the ALICE results are normalized to inelastic events and CMS results are normalized to non-single-diffractive (NSD) events. 
Under the assumption that the multi-strange baryon production at mid-rapidity is negligible in single-diffractive events, the difference is solely due to 
the cross-section ratio $\sigma_{\rm NSD}/\sigma_{\rm inelastic}$, which the comparison shows to be $76\pm13\%$. The \meanpt\ measured by ALICE and CMS for $\Xi^{\pm}$ is in agreement.

\begin{figure}
\centerline{\epsfxsize 3.35in \epsffile{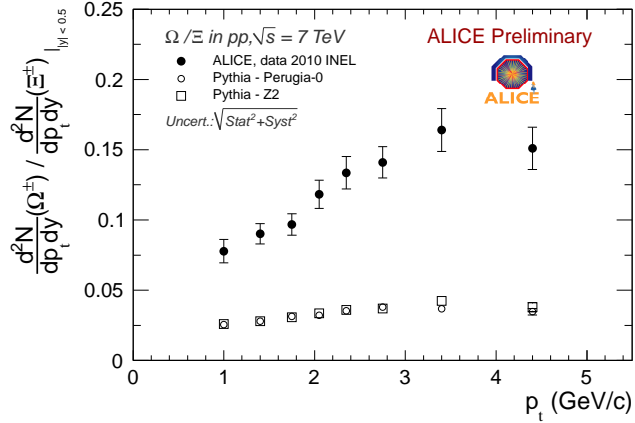}}
\caption{\label{SecondFigure} The ratio of $\Xi^{\pm}$ and $\Omega^{\pm}$ transverse momentum spectra compared to MC predictions.}
\end{figure}

In order to compute the ratio of $\Omega^{\pm}/\Xi^{\pm}$ as a function of \ptee, the $\Xi^{\pm}$ analysis was also performed in the same transverse momentum binning as the $\Omega^{\pm}$. 
The resulting ratio can be seen in Fig. \ref{SecondFigure} and is shown to rise with momentum, which is consistent with the different slopes 
observed in Fig. \ref{ActualSpectra}. PYTHIA predictions are also shown and do not reproduce this 
ratio well in any of the tunes used. This information could be used to further tune PYTHIA to reproduce the strangeness production rates seen in \pp\ collisions at LHC energies. 

\section{Status of the analysis of Pb-Pb data}

\begin{figure}
\centerline{\epsfxsize 1.85in \epsffile{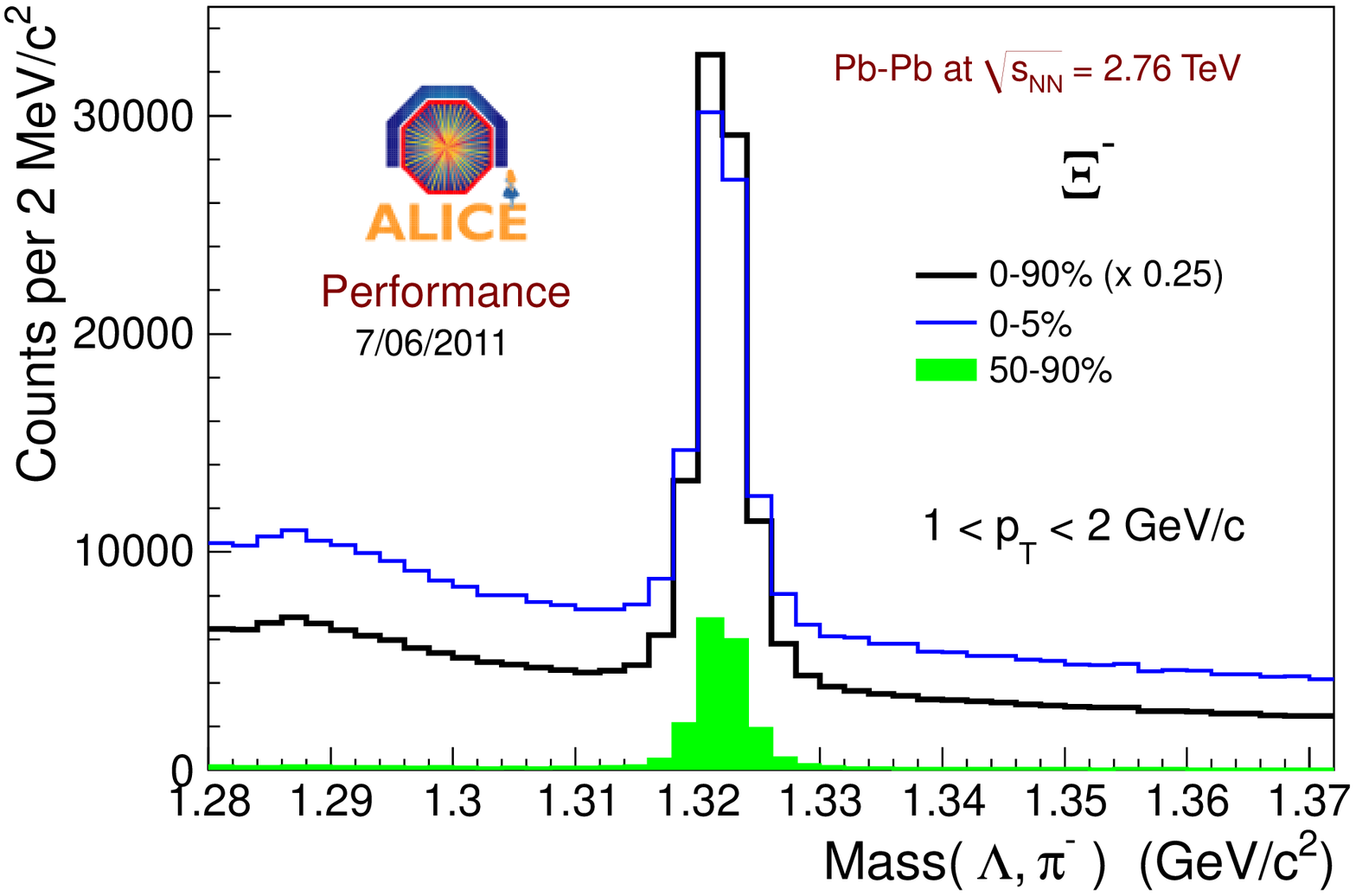} \epsfxsize 1.85in \epsffile{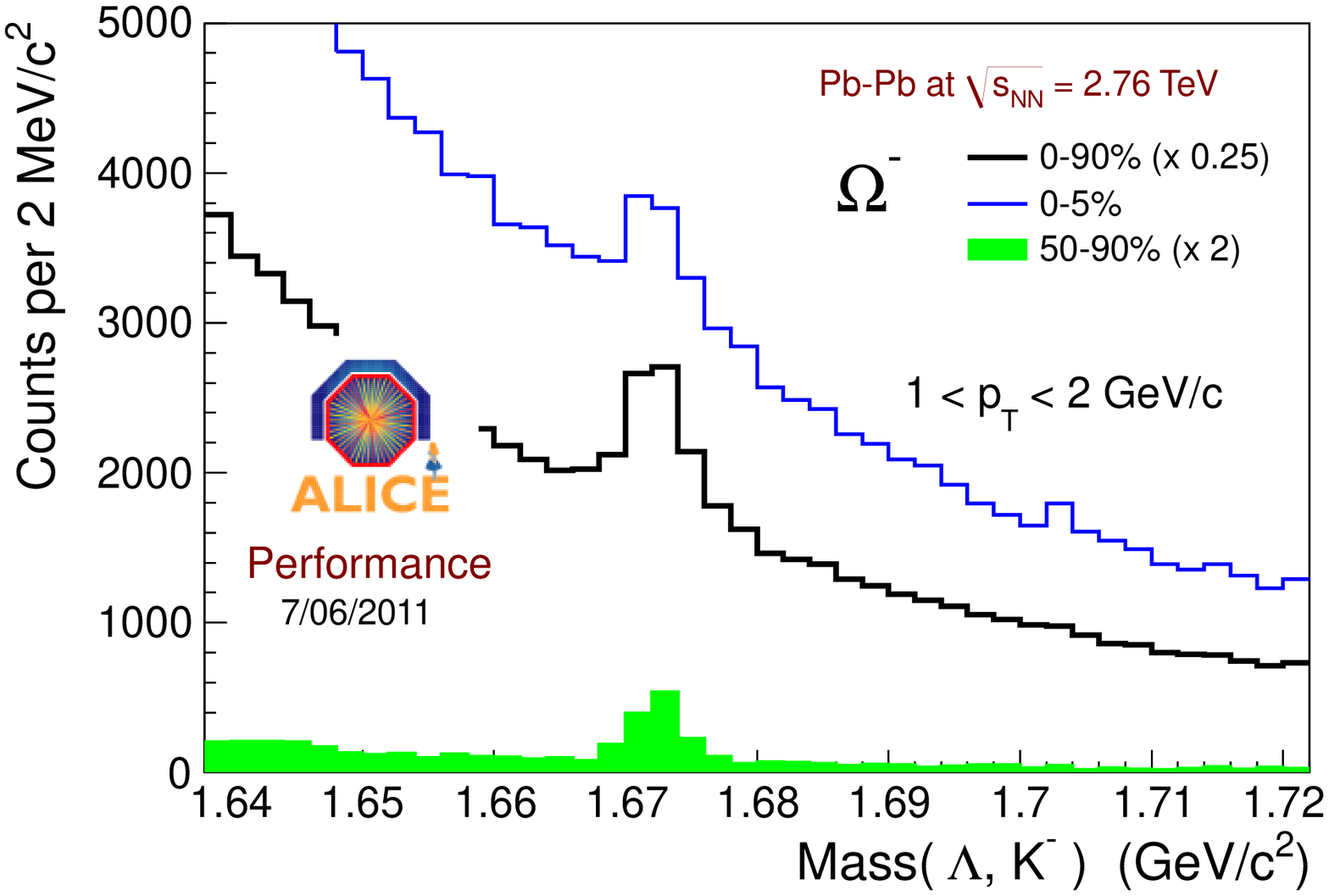}}
\centerline{\epsfxsize 1.85in \epsffile{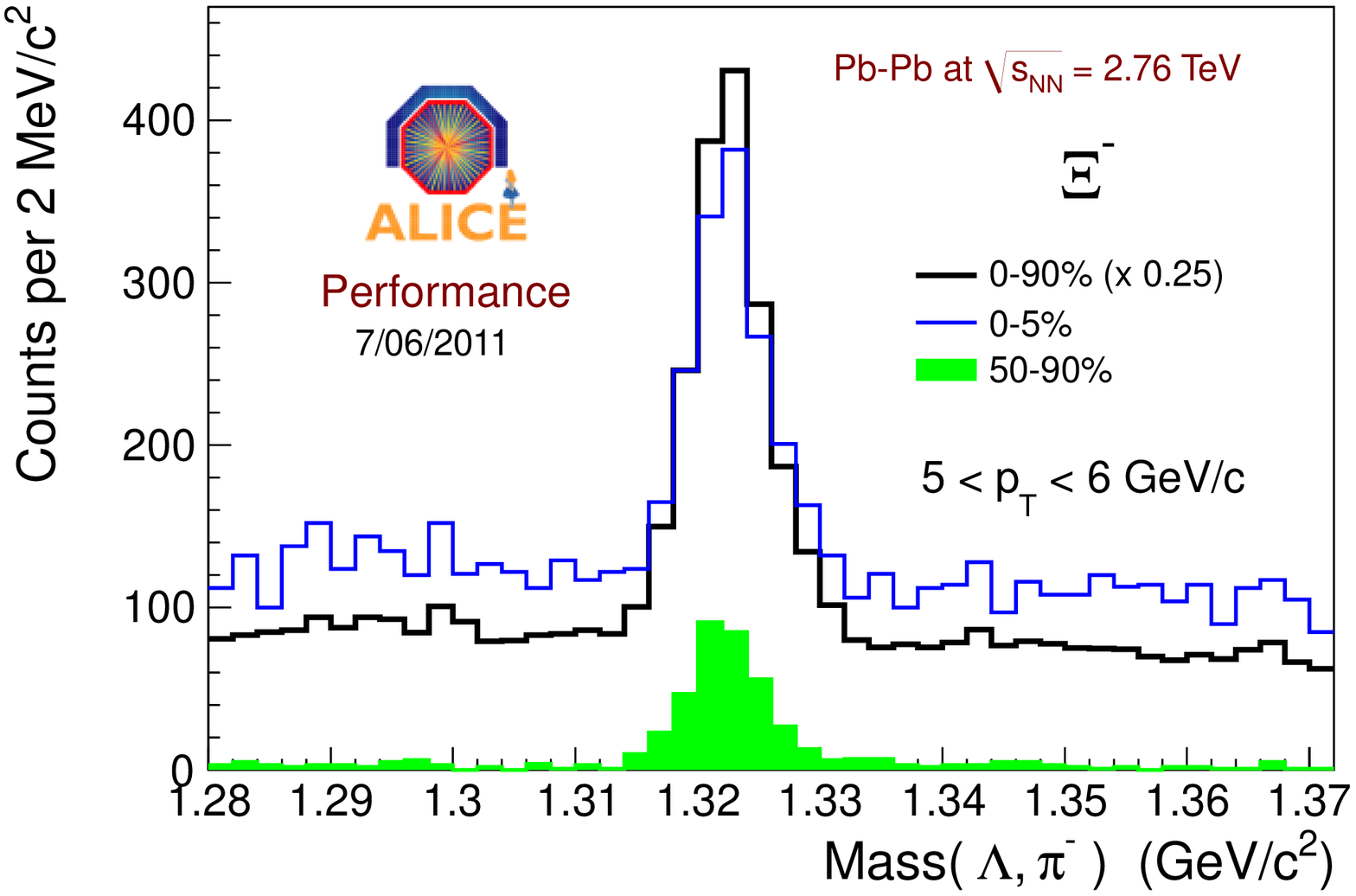} \epsfxsize 1.85in \epsffile{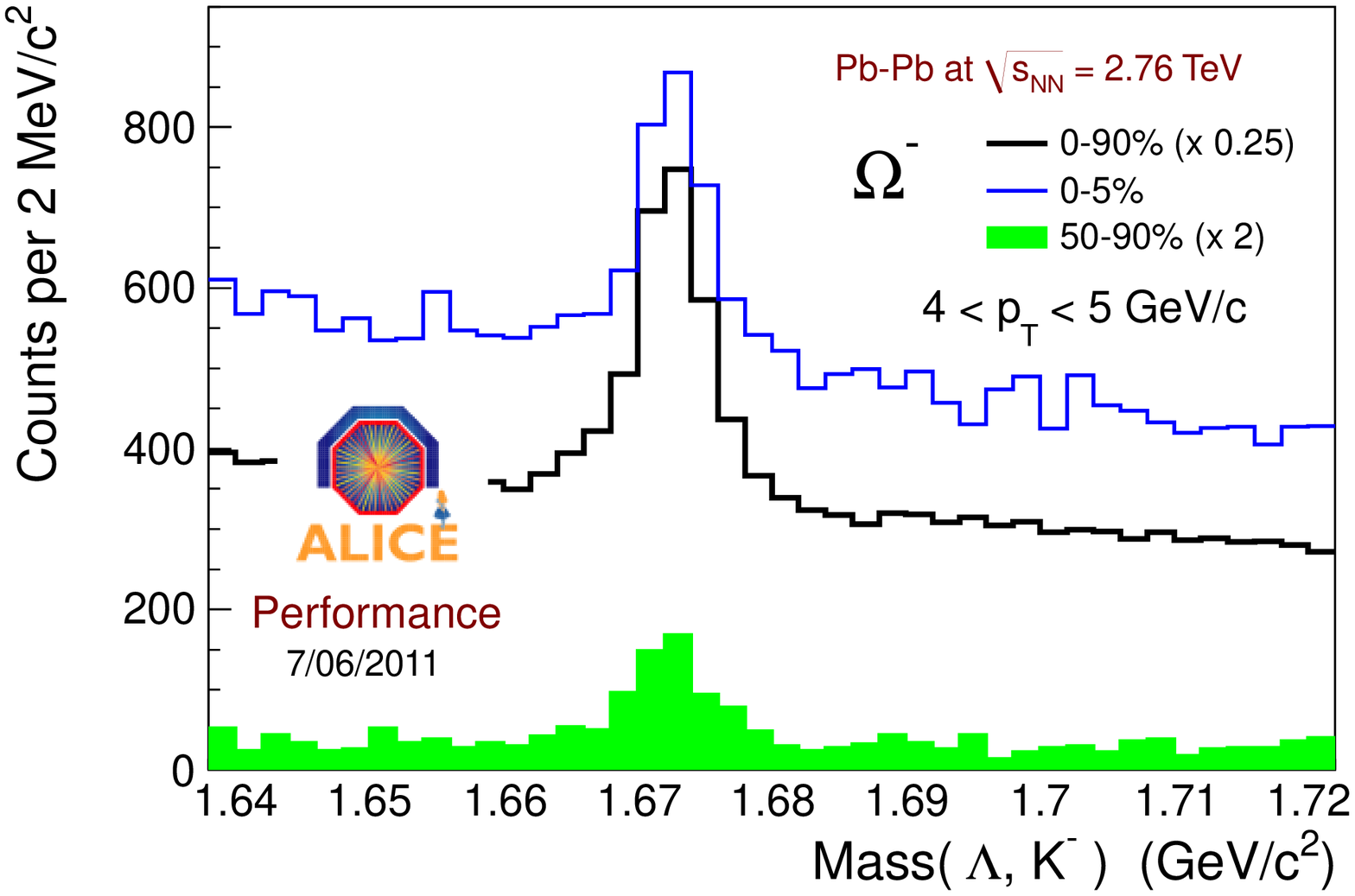}}
\caption{\label{PbPbPerformance} $\Xi^{\pm}$ and $\Omega^{\pm}$ invariant mass distributions in peripheral and central Pb-Pb collisions and in high and low transverse momentum bins. The minimum bias distributions are shown as black histograms.}
\end{figure}

Using a data sample of $15\times10^6$ lead-lead (Pb-Pb) collisions as measured by ALICE in 2010, we show in Fig. \ref{PbPbPerformance} the invariant mass distributions for the multi-strange baryons.
A clear signal can be seen in both a wide transverse momentum range (from $1.0-6.0$ GeV/$c$ for the $\Xi^{-}$ and \Ix\ and $1.0-5.0$ GeV/$c$ for the $\Omega^{-}$ and \Mo) and a large reach
in centrality, making it possible to perform spectrum extraction even in peripheral event categories such as 60-80$\%$. The combinatorial background is significantly larger 
than in \pp\ collisions, so that topological cuts had to be tightened to improve the signal to background ratio. Comparisons to pp systems will be drawn once the proper efficiency corrections have been applied. 

\section*{Acknowledgements} 
We wish to thank Funda\c{c}\~ao de Amparo a Pesquisa do Estado de S\~ao Paulo, FAPESP,
Brazil for the support.

\section*{References}

\bibliographystyle{jphysg}
\bibliography{proc_v23}


\end{document}